\documentclass[proceedings]{JHEP3} 

\usepackage{epsfig}                   

\PrHEP{}   
\conference{International Workshop on Astroparticle and High Energy Physics}

\title{Double beta decay and the extra-dimensional seesaw mechanism}

\author{
Gautam Bhattacharyya\\
Saha Institute of Nuclear Physics, 1/AF Bidhan Nagar,
         Kolkata 700064, India\\
        E-mail: \email{gb@theory.saha.ernet.in}}

\author{
Hans Volker Klapdor--Kleingrothaus\\
Max--Planck--Institut f\"ur Kernphysik, P.O. Box 103980, D--69029 Heidelberg,
Germany\\
        E-mail: \email{klapdor@gustav.mpi-hd.mpg.de}}

\author{\speaker{
Heinrich P\"as}
\\
Institut f\"ur Theoretische Physik und Astrophysik,
Universit\"at W\"urzburg,\\ D-97074 W\"urzburg, Germany\\
E-mail: \email{paes@physik.uni-wuerzburg.de}}

\author{
Apostolos Pilaftsis\\
Department of Physics and Astronomy, University of Manchester,\\ 
         Manchester M13 9PL, United Kingdom\\
        E-mail: \email{pilaftsi@a35.ph.man.ac.uk}}

\abstract{
We  study  the  model-building   conditions  under  which  an  
observable $0\nu\beta\beta$-decay signal 
is predicted
due to Kaluza--Klein singlet neutrinos in theories with large extra
dimensions.  Our  analysis is based  on 5-dimensional singlet-neutrino
models   compactified   on    an   $S^1/Z_2$   orbifold,   where   the
Standard--Model fields  are localized  on a 3-brane.   We show  that 
sizeable
$0\nu\beta\beta$ rates within  the  above
minimal 5-dimensional framework would require a non-vanishing shift of
the 3-brane from  the orbifold fixed points by  an amount smaller than
the  typical scale (100~MeV)$^{-1}$  characterizing the  Fermi nuclear
momentum.   The  resulting  5-dimensional  models predict  a  sizeable
effective  Majorana-neutrino  mass that  could  be  several orders  of
magnitude larger than the  light neutrino masses.  
}

\begin{document}

\section{Introduction}

Recently,   realizations of  phenomenologically  viable  theories with
large   compact  dimensions   of~TeV size~\cite{extra}  have  enriched
dramatically  our perspectives in  searching   for physics beyond  the
Standard  Model   (SM).     Among   the  possible   higher-dimensional
realizations,  sterile    neutrinos  propagating   in     large  extra
dimensions~\cite{DDG2,ADDM,AP1,IP}    may  provide
interesting  alternatives for generating  the observed  light neutrino
masses.  On the other hand, detailed  experimental studies of neutrino
properties may even shed light on the geometry and/or shape of the new
dimensions.  In this context, one  of the most sensitive  experimental
approaches to neutrino  masses and their  properties is the search for
neutrinoless double  beta decay~\cite{doi85}. Neutrinoless double beta
decay, denoted in short as $0\nu\beta\beta$, corresponds to two single
beta  decays~\cite{Klapdor}    occurring simultaneously    in  one
nucleus,  thereby  converting  a    nucleus  $(Z,A)$  into  a  nucleus
$(Z+2,A)$, i.e.\
\begin{displaymath}        
^{A}_{Z}\,X\ \to\ ^A_{Z+2}\,X\: +\: 2 e^-\; . 
\end{displaymath}
This process   violates  lepton number by   two  units  and  hence its
observation would  signal physics  beyond  the  SM.  To a   very  good
approximation, the half life  for a $0\nu\beta\beta$ decay mediated by
light neutrinos is given by
\begin{equation}
\label{t1/2}
[T^{0\nu\beta\beta}_{1/2}]^{-1}\ =\ \frac{|\langle m \rangle |^2}{m^2_e}\
|{\cal M}_{0\nu\beta\beta}|^2\, G_{01}\; , 
\end{equation} 
where $\langle  m  \rangle$ denotes  the effective  neutrino  Majorana
mass, $m_e$  is the electron mass  and ${\cal M}_{0\nu\beta\beta}$ and
$G_{01}$ denote the appropriate 
nuclear matrix element and the phase space factor,
respectively.   For    details, see~\cite{doi85,Klapdor}  and  our
discussion in \cite{Bhattacharyya:2002vf}.
 
An analysis of the Heidelberg-Moscow experiment 
reports an evidence for
$|\langle m \rangle|=
0.39^{+0.45}_{-0.34}~{\rm eV}\quad (95\%~{\rm CL})$ \cite{HMexp}.
The IGEX experiment derives a bound of
$|\langle m \rangle| < 0.33-1.35~{\rm eV}\quad (95\%~{\rm CL})$~\cite{igex}.
The NEMO3 experiment, which may come to a sensitivity in this range,
has started operation \cite{nemo}.
Several proposals for next-generation experiments
aim at sensitivities down to (few)~$\times 10^{-2}$~eV, among these the
GENIUS, CUORE and EXO projects with approved prototypes (see \cite{elliott} 
for an overview).

Here, we focus on the question, whether extra dimensional neutrino models
can be tested by double beta experiments via an observable signal
in present or next generation setups.
Within the  framework of  theories   with large   extra dimensions,  previous
studies on neutrinoless  double beta decays were performed  within the
context  of   higher-dimensional models   that  utilize  the   shining
mechanism from a distant brane~\cite{MLP} and of theories with wrapped
geometric       space~\cite{Huber}.         In   Ref.~\cite{MLP},  the
$0\nu\beta\beta$ decay  is   accompanied with emission    of Majorons,
whereas the prediction in~\cite{Huber}  falls  short by two  orders of
magnitude to account for an observable signal in running experiments.

Here we   consider an even more  minimal  higher-dimensional
framework of lepton-number violation, namely 5-dimensional theories 
compactified  on a $S^1/Z_2$ orbifold,
in  which only one 5-dimensional
(bulk) sterile neutrino is added  to the field content of  the SM
\cite{Bhattacharyya:2002vf}.   
In this minimal model, the SM   fields are localized on a   4-dimensional
Minkowski subspace, also termed 3-brane. The model naturally generates 
small neutrino masses in an higher-dimensional analogue of the seesaw 
mechanism \cite{DDG2}.

\section{Minimal higher-dimensional neutrino models}

\FIGURE[t]{\epsfig{file=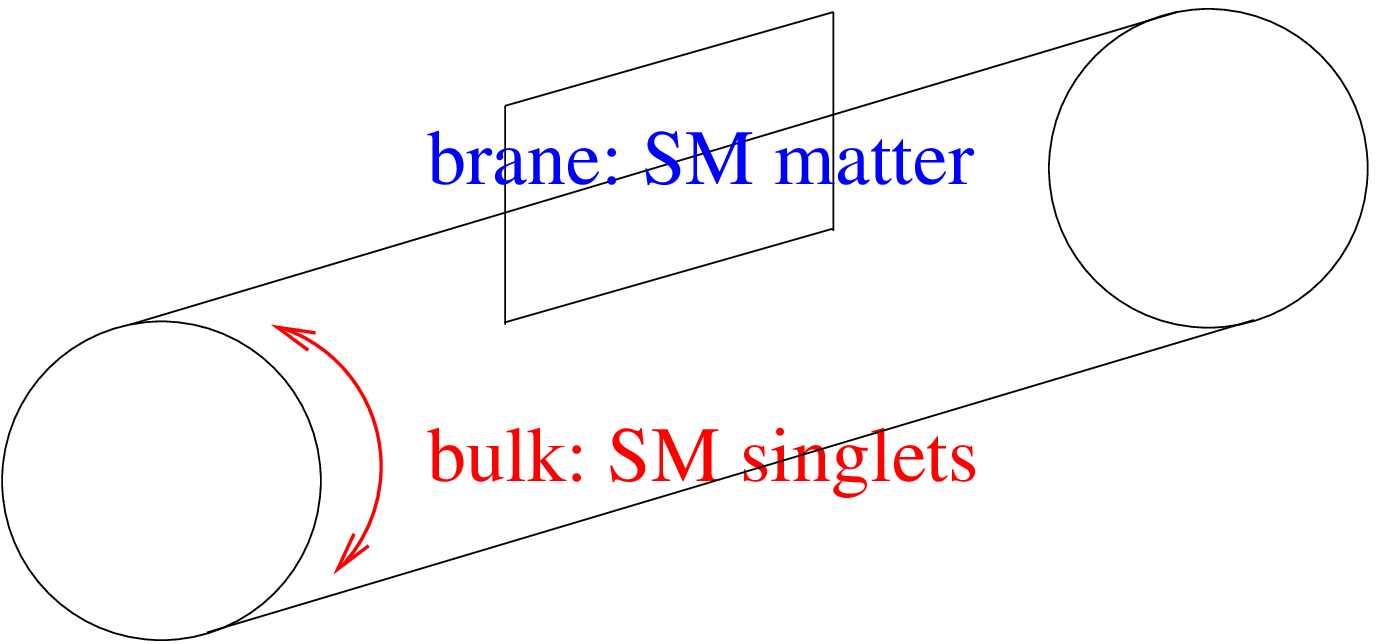, width=.8\textwidth}%
	     \caption{The SM matter is localized on a 3-brane, while
the sterile singlet neutrinos are allowed to propagate in the bulk.
This frameowork naturally generates small neutrino masses.}
\label{extradim}}	

In  this section, we will describe  the  basic low-energy structure of
minimal higher-dimensional extensions of the  SM that include  singlet
neutrinos.   In  particular, we assume    that singlet neutrinos being
neutral under   the SU(2)$_L\otimes$U(1)$_Y$ gauge  group  can  freely
propagate   in  a    higher-dimensional   space  of   $[1+(3+\delta)]$
dimensions, the so-called bulk, whereas all SM particles are localized
in  a $(1+3)$-dimensional subspace, known  as 3-brane or simply brane.
However,  even singlet neutrinos themselves may  live in a subspace of
an even  higher-dimensional  space of $[1+(3+n_g)]$  dimensions,  with
$\delta \le n_g$, in which gravity propagates.

We shall restrict  our study to 5-dimensional  models, i.e.\  the case
$\delta  =  1$,  where the singlet   neutrinos  are compactified on  a
$S^1/Z_2$  orbifold.   Specifically,  the   leptonic  sector  of   our
5-dimensional model consists of the SM lepton fields:
\begin{equation}
  \label{LSM}
L(x)\ =\ \left( \begin{array}{c} \nu_l (x) \\ l_L (x) \end{array}
\right) ,\qquad
l_R (x)\,,
\end{equation}
with $l = e,\mu,\tau$, and one 5-dimensional (bulk) singlet neutrino:
\begin{equation}
  \label{Nu}
 N(x,y)\ =\ \left( \begin{array}{c} \xi (x,y) \\ 
\bar{\eta} (x,y) \end{array} \right)\, ,
\end{equation}
where  $y$ denotes  the  additional compact dimension,  and $\xi$  and
$\eta$ are 5-dimensional two-component   spinors.  
For generality, we will assume that the brane,
where the SM leptons are localized,
is shifted from the orbifold fixed point $y = 0$ to $y = a$.

As   usual,  we  impose the   periodic   boundary condition $N(x,y)  =
N(x,y+2\pi R)$ with  respect to $y$ dimension  on the singlet neutrino
field.  In  addition,  the  action  of  $S^1/Z_2$ orbifolding  on  the
5-dimensional spinors  $\xi$    and $\eta$   entails  the   additional
identifications:
\begin{equation}
  \label{yparity}
\xi (x,y)\ =\ \xi (x,-y)\,,\qquad  \eta (x,y)\ =\ - \eta (x,-y)\,.
\end{equation}
In other   words,  the spinors  $\xi$ and   $\eta$ are   symmetric and
antisymmetric under a $y$ reflection, respectively.

With the above  definitions, the most generic  effective 4-dimensional
Lagrangian       of        such      a      model          is    given
by~\cite{DDG2,AP1}.
\begin{eqnarray}
  \label{Leff} 
{\cal L}_{\rm  eff} & =& \int\limits_0^{2\pi R}\!\! dy\
\bigg\{\,   \bar{N} \Big(   i\gamma^\mu  \partial_\mu\,  +\,  \gamma_5
\partial_y \Big) N\ -\  \frac{1}{2}\,\Big( M N^T  C^{(5)-1} N\ +\ {\rm
h.c.}    \Big)         \nonumber\\     
&&+\,\delta   (y-a)\,  \bigg[\, \frac{{h}^l_1}{(M_F)^{\delta/2}}\, 
L\tilde{\Phi}^*      \xi\,    +\, \frac{{h}^l_2}{(M_F)^{\delta/2}}\,  L 
\tilde{\Phi}^*   \eta\ +\   {\rm h.c.}\,\bigg]\ +\ 
\delta (y-a)\, {\cal L}_{\rm SM}\, \bigg\}\, ,\quad
\end{eqnarray}
where $\tilde{\Phi} =  i\sigma_2 \Phi^*$ is the  hypercharge-conjugate
of the SM Higgs  doublet $\Phi$, with hypercharge  $Y(\Phi) = 1$,  and
${\cal L}_{\rm SM}$ denotes the SM Lagrangian which is restricted on a
brane at $y=a$~\cite{DDG2}.  In   addition, $M_F$ is   the fundamental
$n_g$-dimensional Planck scale and $\delta =  1$ for sterile neutrinos
propagating  in 5 dimensions.  Notice  that the mass term $m_D \bar{N}
N$ is not allowed  in~(\ref{Leff}), as a result  of the $Z_2$ discrete
symmetry.  

We now proceed with  the compactification of  the $y$ dimension of the
$S^1/Z_2$     orbifold  model.  Because    of   their   symmetric  and
antisymmetric   properties~(\ref{yparity}) under   $y$ reflection, the
two-component spinors $\xi$ and $\eta$   can be expanded in a  Fourier
series of cosine and sine harmonics:
\begin{eqnarray}
  \label{xi}
\xi (x,y) &=& \frac{1}{\sqrt{2\pi R}}\ \xi_0 (x)\ +\ 
\frac{1}{\sqrt{\pi R}}\ \sum_{n=1}^\infty\, \xi_n (x)\ 
                                       \cos\bigg(\,\frac{ny}{R}\,\bigg)\,,\\
  \label{eta}
\eta (x,y) & =& \frac{1}{\sqrt{\pi R}}\ \sum_{n=1}^\infty\, \eta_n (x)\ 
                                       \sin\bigg(\,\frac{ny}{R}\,\bigg)\,,
\end{eqnarray}
where the chiral spinors $\xi_n (x)$ and $\eta_n (x)$ form an infinite
tower of   KK  modes. 

After substituting~(\ref{xi})  into  (\ref{Leff}) and  integrating out
the $y$ coordinate, we obtain the effective 4-dimensional Lagrangian
\begin{eqnarray}
  \label{Leff1KK}
{\cal L}_{\rm eff} & = & {\cal L}_{\rm SM}\ +\ \bar{\xi}_0
( i\bar{\sigma}^\mu \partial_\mu) \xi_0\ 
+\ \Big(\, \bar{h}^{l(0)}_1\, L\tilde{\Phi}^* \xi_0\ -\
\frac{1}{2}\, M\, \xi_0\xi_0\ +\ {\rm h.c.}\,\Big)\
 +\ \sum_{n=1}^\infty\, \bigg[\, \bar{\xi}_n 
( i\bar{\sigma}^\mu \partial_\mu) \xi_n\nonumber\\
&& +\, \bar{\eta}_n ( i\bar{\sigma}^\mu \partial_\mu) \eta_n\
+\ \frac{n}{R}\, \Big( \xi_n \eta_n\, +\, \bar{\xi}_n
\bar{\eta}_n\Big) -\ \frac{1}{2}\, M\, 
\Big( \xi_n\xi_n\, +\, \bar{\eta}_n\bar{\eta}_n\
+\ {\rm h.c.}\Big)\nonumber\\
&& +\, \sqrt{2}\, \Big(\, \bar{h}^{l(n)}_1\, L\tilde{\Phi}^* \xi_n\ +\
\bar{h}^{l(n)}_2\, L\tilde{\Phi}^* \eta_n\ +\ {\rm h.c.}\,\Big)\, \bigg]\, ,
\end{eqnarray}
where 
\begin{eqnarray}
  \label{h1n}
\bar{h}^{l(n)}_1 &=& \frac{h^l_1}{(2\pi M_F R)^{\delta/2}}\ 
\cos \bigg(\,\frac{na}{R}\,\bigg)\ =\ \bigg(\,\frac{M_F}{M_{\rm
P}}\,\bigg)^{\delta/n_g}\  
h^l_1 \cos \bigg(\,\frac{na}{R}\,\bigg)\,,\\
  \label{h2n}
\bar{h}^{l(n)}_2 &=& \frac{h^l_2}{(2\pi M_F R)^{\delta/2}}\ 
\sin \bigg(\,\frac{na}{R}\,\bigg)\ =\ \bigg(\,\frac{M_F}{M_{\rm
P}}\,\bigg)^{\delta/n_g}\  
h^l_2 \sin \bigg(\,\frac{na}{R}\,\bigg)\, . 
\end{eqnarray}
In deriving the last step on the RHS's of (\ref{h1n}) and (\ref{h2n}),
we have employed the basic relation among the Planck mass $M_{\rm P}$,
the corresponding $n_g$-dimensional     Planck mass   $M_F$  and   the
compactification radii $R$ (all taken to be of equal size):
\begin{equation}
  \label{MF}
M_{\rm P}\ =\ (2\pi\, M_F\, R)^{n_g/2}\, M_F\ .
\end{equation}
{}From~(\ref{h1n})   and  (\ref{h2n}),  we    see  that  the   reduced
4-dimensional Yukawa couplings $\bar{h}^{(n)}_{1,2}$ can be suppressed
by   many orders of  magnitude\cite{ADDM,DDG2}  if  there  is  a large
hierarchy between $M_{\rm  P}$  and the quantum  gravity  scale $M_F$.
Thus, if  gravity  and bulk neutrinos feel  the  same number of  extra
dimensions, i.e.\ $\delta  = n_g$, the 4-dimensional  Yukawa couplings
$\bar{h}^{(n)}_1$ and $\bar{h}^{(n)}_2$  are naturally suppressed by a
huge  factor $M_F/M_{\rm P} \sim 10^{-15}$,  for $M_F \approx 10$~TeV.
{}From~(\ref{Nu}), we  observe  that $\xi$ and $\bar{\eta}$  belong to
the same multiplet   and hence have   the same lepton number. It  then
follows   from  (\ref{Leff1KK})  that  the  simultaneous  presence  of
$\bar{h}^{(n)}_1$ and $\bar{h}^{(n)}_2$ in  an amplitude gives rise to
lepton number violation by two units.

If the brane were located at the one of the two orbifold fixed points,
e.g.\ at $y=0$, the operator  $L\tilde{\Phi}^*\eta$ would be absent as
a consequence  of the $Z_2$ discrete symmetry.   However, if the brane
is shifted  by an amount   $a\ne 0$, the  above  operator is no longer
absent. In fact, the coexistence of the
two    operators $L\tilde{\Phi}^*\xi$ and $L\tilde{\Phi}^*\eta$ breaks
the  lepton  number leading  to   observable effects  in  neutrinoless
double beta decay experiments.

Let us now introduce the weak basis for the KK-Weyl spinors
\begin{equation}
  \label{xieta}
\chi_{\pm n}\ =\ \frac{1}{\sqrt{2}}\, (\,\xi_n\: \pm\:
\eta_n\,).
\end{equation}

Following~\cite{DDG2},    we rearrange  the   singlet KK-Weyl  spinors
$\xi_0$ and $\chi^\pm_n$, such that the smallest diagonal entry of the
KK neutrino  mass  matrix  is
$|\varepsilon|  = {\rm min}\,  \Big(  | M -  \frac{k}{R}  | \Big) \leq
1/(2R)$, for a given value $k=k_0$.   In this newly defined basis, the
effective kinetic Lagrangian becomes
\begin{equation}
  \label{Lkinorb}
{\cal L}_{\rm kin}\ =\ \frac{1}{2}\, \bar{\Psi}_\nu\,\Big(\, 
i\!\not\!\partial\ -\ {\cal M}^{\rm KK}_\nu\, \Big)\, \Psi_\nu\,,
\end{equation}
where $\Psi_\nu$ is the reordered (4-component) Majorana-spinor vector
\begin{equation}
  \label{Psinu}
\Psi^T_\nu \ =\ \left[\, 
\left(\! \begin{array}{c} \nu_l \\ \bar{\nu}_l \end{array}\!\right)\,,\
\left(\! \begin{array}{c} \chi_{k_0} \\ \bar{\chi}_{k_0} 
                                               \end{array}\!\right)\,,\
\left(\! \begin{array}{c} \chi_{k_0+1} \\ \bar{\chi}_{k_0+1} 
                                               \end{array}\!\right)\,,\
\left(\! \begin{array}{c} \chi_{k_0-1} \\ \bar{\chi}_{k_0-1} 
                                               \end{array}\!\right)\,,\
\cdots\,,
\left(\! \begin{array}{c} \chi_{k_0+n} \\ \bar{\chi}_{k_0+n} 
                                               \end{array}\!\right)\,,\
\left(\! \begin{array}{c} \chi_{k_0-n} \\ \bar{\chi}_{k_0-n} 
                                               \end{array}\!\right)\,,\
\cdots\ \right]
\end{equation}
and ${\cal M}^{\rm KK}_\nu$ the corresponding KK neutrino mass matrix
\begin{equation}
  \label{Morb}
{\cal M}^{\rm KK}_\nu\ =\ \left(\! \begin{array}{ccccccc}
0 & m & m & m & m & m & \cdots \\
m & \varepsilon & 0 & 0 & 0 & 0 & \cdots  \\
m & 0 & \varepsilon + \frac{1}{R} & 0 & 0 & 0 & \cdots \\
m & 0 & 0 & \varepsilon - \frac{1}{R} & 0 & 0 & \cdots \\
m & 0 & 0 & 0 & \varepsilon + \frac{2}{R} & 0 & \cdots \\
m & 0 & 0 & 0 & 0 & \varepsilon - \frac{2}{R} & \cdots \\
\vdots & \vdots & \vdots & \vdots & \vdots & \vdots & \ddots
\end{array}\!\right),
\end{equation}
with $m = v\bar{h}_1/\sqrt{2}$.  In  a three-generation model, $m$ and
$\bar{h}_1$ are both 3-vectors in  the flavour space, i.e.\ $\bar{h}_1
=  (\bar{h}^e_1,\ \bar{h}^\mu_1,\  \bar{h}^\tau_1)^T$. For simplicity
we assume here that $\bar{h}_1 = \bar{h}^e_1$.

The  eigenvalues of ${\cal M}^{\rm  KK}_\nu$ can  be computed from the
characteristic eigenvalue  equation  ${\rm   det}\, ( {\cal    M}^{\rm
KK}_\nu - \lambda {\bf 1}) = 0$, which is analytically given by
\begin{equation}
  \label{detorb} 
\prod\limits_{n=0}^\infty\,\bigg[ \Big( \lambda\,
-\,\varepsilon\Big)^2\: -\: \frac{n^2}{R^2}\,\bigg]\,\bigg[\,
1\: +\: \frac{\varepsilon}{\lambda\,
-\,\varepsilon}\ -\ m^2\, \sum\limits_{n=-\infty}^\infty\,\frac{1}{
(\lambda\,-\,\varepsilon)^2\: -\: \frac{n^2}{R^2} }\ \bigg]\ =\ 0\, .
\end{equation}
Since it can be shown that $\lambda - \varepsilon =  \pm n/R$ is never
an  exact  solution to  the  characteristic equation,  only the second
factor in  (\ref{detorb})  can  vanish.   Employing  complex   contour
integration   techniques, the   summation  in the    second factor  in
(\ref{detorb}) can be  performed   exactly, leading to an   equivalent
transcendental equation
\begin{equation}
  \label{transI}
\lambda\ =\ \pi m^2 R\ 
          \cot\, \Big[\,\pi R\, (\lambda - \varepsilon )\,\Big]\, .
\end{equation}
As  was  already discussed  in   \cite{DDG2}, if  $\varepsilon   = 0$,
(\ref{transI})  implies that the mass  spectrum consists of massive KK
Majorana neutrinos degenerate in pairs with opposite  CP parities.  If
$\varepsilon =  1/(2R)$, the  KK  mass  spectrum contains  a  massless
state,  which is predominantly  left-handed   if $mR  < 1$, while  the
remaining massive  KK states  form  degenerate pairs with  opposite CP
parities,  exactly  as in the   $\varepsilon  = 0$ case.   However, if
$\varepsilon   \neq   0,\     1/(2R)$,  the   lepton      number  gets
broken.\footnote{Alternatively, lepton  number may  also be broken
through   the    Scherk-Schwarz  mechanism \cite{SS}, 
where   the Scherk-Schwarz
rotation  angle will induce  terms   very similar  to those  depending
on~$\varepsilon$~\cite{DDG2,DGQ}.}  In this case, there is no massless
state in   the spectrum,  and the  above   exact degeneracy  among the
massive Majorana   neutrinos  becomes only  approximate, with  a  mass
splitting of   order $2\varepsilon$  for each  would-be  ($\varepsilon
\rightarrow 0$) degenerate KK pair.

We now consider   an  orbifold model,   in  which the $y=0$ brane   is
displaced  from the orbifold  fixed points  by an  amount  $a$.  Under
certain  restrictions in Type I  string theory~\cite{GP,DDG2}, such an
operation can be  performed  respecting the  $Z_2$  invariance  of the
original higher-dimensional action.    In  particular, one   can  take
explicitly account of this last property  by considering the following
replacements in the effective Lagrangian (\ref{Leff}):
\begin{eqnarray}
\xi\, \delta (y - a) &\to & \frac{1}{2}\; \xi\,\Big[\, \delta ( y - a)\: +\:
\delta ( y + a - 2\pi R) \, \Big]\,,\nonumber\\
\eta\, \delta (y - a) &\to & \frac{1}{2}\; \eta\,\Big[\, \delta ( y - a)\: -\:
\delta ( y + a - 2\pi R) \, \Big]\,,
\end{eqnarray}
with $0\le a < \pi R$ and $0\le  y \le 2\pi R$.  It  is obvious that a
$Z_2$-invariant implementation of brane-shifted couplings requires the
existence of two branes at least, placed  at $y= a$ and $y  = 2\pi R -
a$. 

Proceeding  as  above, the effective   KK neutrino mass  matrix ${\cal
M}^{\rm KK}_\nu$ for  the orbifold model  with a shifted brane can  be
written down in an analogous form
\begin{equation}
  \label{Morbshift}
{\cal M}^{\rm KK}_\nu\ =\ \left(\! \begin{array}{ccccccc}
0 & m^{(0)} & m^{(1)} & m^{(-1)} & m^{(2)} & m^{(-2)} & \cdots \\
m^{(0)} & \varepsilon & 0 & 0 & 0 & 0 & \cdots  \\
m^{(1)} & 0 & \varepsilon + \frac{1}{R} & 0 & 0 & 0 & \cdots \\
m^{(-1)} & 0 & 0 & \varepsilon - \frac{1}{R} & 0 & 0 & \cdots \\
m^{(2)} & 0 & 0 & 0 & \varepsilon + \frac{2}{R} & 0 & \cdots \\
m^{(-2)} & 0 & 0 & 0 & 0 & \varepsilon - \frac{2}{R} & \cdots \\
\vdots & \vdots & \vdots & \vdots & \vdots & \vdots & \ddots
\end{array}\!\right)\,,
\end{equation}
where 
\begin{eqnarray}
  \label{mk0}
m^{(n)} &=& \frac{v}{\sqrt{2}}\, \bigg[\, \bar{h}_1\, 
\cos\bigg(\frac{ (n-k_0) a}{R}\bigg)\: +\: \bar{h}_2\, 
\sin\bigg(\frac{ (n-k_0) a}{R}\bigg)\,\bigg]\ =\
m\,\cos\bigg(\frac{na}{R}\, -\, \phi_h\,\bigg)\,,\qquad
\end{eqnarray}
with $m  =  v  \sqrt{(\bar{h}^2_1  + \bar{h}^2_2)/2}$   and  $\phi_h =
\tan^{-1} (\bar{h}_2 / \bar{h}_1) + k_0 a /R$.  As before, we consider
an one-generation model  with $\bar{h}_1 = \bar{h}^e_1$ and $\bar{h}_2
= \bar{h}^e_2$,  which    renders the analytic   determination of  the
eigenvalue   equation  tractable.  
Thus, for our one-generation brane-shifted
model, the characteristic eigenvalue equation reads
\begin{equation}
  \label{detorbshift} 
\prod\limits_{n=0}^\infty\,\bigg[ \Big( \lambda\,
-\,\varepsilon\Big)^2\: -\: \frac{n^2}{R^2}\,\bigg]\,\bigg[\,
1\: +\: \frac{\varepsilon}{\lambda\,
-\,\varepsilon}\ -\ \frac{1}{\lambda - \varepsilon}\,
\sum\limits_{n=-\infty}^\infty\,\frac{m^{(n)\, 2}}{
\lambda\: -\: \varepsilon\: -\: \frac{n}{R}\:   }\ \bigg]\ =\ 0\, ,
\end{equation}
which is equivalent to
\begin{equation}
  \label{eigen}
\lambda\ =\ \sum\limits_{n=-\infty}^\infty\;
\frac{m^{(n)\, 2}}{\lambda\: -\: \varepsilon\: -\: \frac{n}{R}}\ .
\end{equation}

We    carry out the   infinite sum  in~(\ref{eigen})
analytically and derive the eigenvalue equation for the simplest class
of cases,  where $a = \pi  R/q$ with $q$  an integer larger
than 1,    i.e.\  $q\ge 2$.    More precisely,   we find
\begin{eqnarray}
  \label{transII}
\lambda &=& \pi m^2 R\ \bigg\{ \cos^2 \Big[\, \phi_h\, -\, a (\lambda
- \varepsilon) \,\Big]\, \cot\Big[\,\pi R\, (\lambda - \varepsilon )\,\Big]
\ -\  \frac{1}{2}\sin \Big[\, 2\phi_h\, -\, 2a (\lambda
- \varepsilon) \,\Big]\ \bigg\}~~~~\, .\qquad
\end{eqnarray}
Observe that unless $\varepsilon = 1/(2R)$, $a =  \pi R/2$ and $\phi_h
= \pi/4$,   the mass spectrum  consists   of massive non-degenerate KK
neutrinos.   However, it can   be shown from~(\ref{transII}) that this
tree-level mass  splitting between a pair of  KK Majorana neutrinos is
generally small for $m_{(n)} \gg 1/R$.  In particular, this tree-level
mass splitting is almost  independent of $a$  and subleading so  as to
play any relevant r\^ole in our calculations.

\section{Effective neutrino-mass estimates}

In this section, we calculate the $0\nu\beta\beta$ observable $\langle
m\rangle$ in orbifold  5-dimensional models.  This quantity determines
the size  of the neutrinoless  double  beta decay amplitude,  which is
induced by $W$-boson exchange graphs.  To this end, it is important to
know the  interactions of the  $W^\pm$  bosons to the  charged leptons
$l=e, \mu,  \tau$  and  the KK-neutrino   mass-eigenstates  $n_{(n)}$.
Adopting the  conventions of~\cite{IP}, the effective charged current
Lagrangian is given by
\begin{equation}
  \label{charged}
{\cal L}^{W^\pm}_{\rm int}\ =\ - \frac{g_w}{\sqrt{2}}\,
W^{-\mu}\, \sum_{l=e,\mu ,\tau}\, \bigg( B_{l\nu_l}\,
\bar{l}\,\gamma_\mu P_L\, \nu_l\: +\: \sum_{n=-\infty}^{+\infty}\,
B_{l,n}\, \bar{l}\,\gamma_\mu P_L\, n_{(n)}\,\bigg)\: +\: {\rm h.c.}\,,
\end{equation}
where $g_w$ is the weak  coupling constant, $P_L = (1-\gamma_5)/2$  is
the  left-handed chirality   projector,    and $B$  is    an  infinite
dimensional mixing matrix.  The   matrix $B$ satisfies  the  following
crucial identities:
\begin{eqnarray}
  \label{B1}
B_{l\nu_l} B^*_{l'\nu_l}\: +\: \sum_{n=-\infty}^{+\infty}\,
B_{l,n} B^*_{l',n} &=& \delta_{ll'}\,,\\
  \label{B2}
B_{l\nu_l}\, m_{\nu_l}\, B_{l'\nu_l}\: +\: \sum_{n=-\infty}^{+\infty}\,
B_{l,n}\, m_{(n)}\, B_{l',n} &=& 0\,.
\end{eqnarray}
Equation (\ref{B1}) reflects  the unitarity properties of  the charged
lepton weak  space,  and (\ref{B2})  holds true, as   a result  of the
absence of the Majorana mass terms $\nu_l \nu_{l'}$ from the effective
Lagrangian in the flavour basis.  For the models under discussion, the
KK neutrino masses $m_{(n)}$     can  be determined exactly  by    the
solutions  of the corresponding   transcendental equations.  To a good
approximation,  however,  these  solutions   for large   $n$  simplify
to
\begin{equation}
  \label{mn}
m_{(n)}\ \approx\  \frac{n}{R}\ + \: \varepsilon\; .
\end{equation}
Like the  neutrino masses, the  mixing-matrix elements
$B_{e\nu}$ and $B_{e,n}$ can also be computed exactly \cite{DDG2}:
\begin{eqnarray}
  \label{Bln} 
B_{e\nu} & = &   \frac{1}{1 + \pi^2 m^2 R^2 +
\frac{m^2_\nu}{m^2}}, \\
B_{e,n} &\simeq&\frac{m^2 \cos^2(\, \frac{na}{R} - \phi_h\,)}
{(\,\frac{n}{R} + \varepsilon\,)^2}.
\label{ben}
\end{eqnarray}
Here we used the
eigenvalue  equation  (\ref{transI}) for  $\lambda  =  m_\nu$.  {}From
(\ref{Bln}), we immediately see that  if $mR \ll 1$
and $m_\nu \ll m$, it is  $B_{e\nu} \approx 1$  and hence the lightest
neutrino state is  predominantly left-handed.
The last  approximate equality in (\ref{ben}) corresponds to
a large  $n$.

\FIGURE[!t]{\epsfig{file=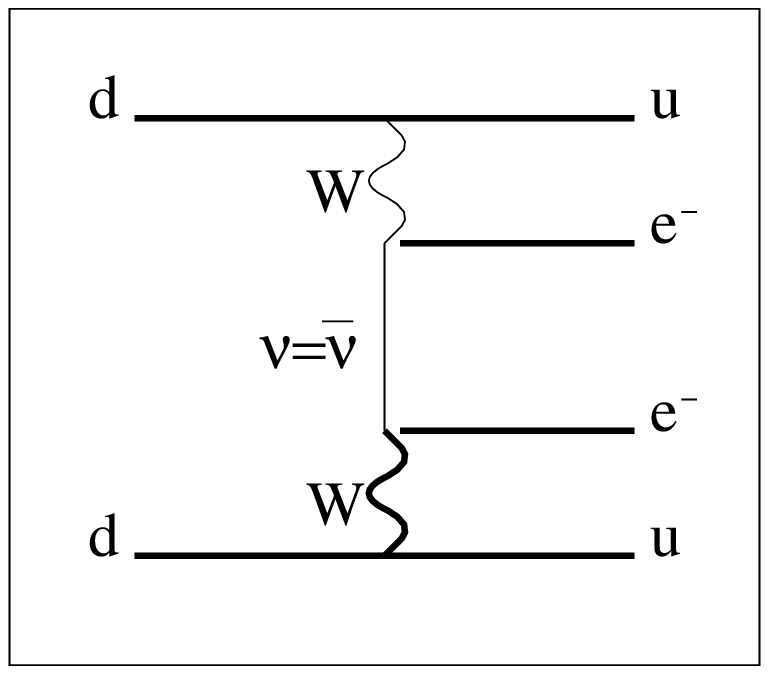, width=.4\textwidth}%
	     \caption{Feynman diagram for neutrinoless double beta decay.
In the extra-dimensional framework, all KK states contribute to the
propagator, weighted with the appropriate mixing and nuclear matrix 
elements.
}\label{0vBB}}	

According to~(\ref{t1/2}), the $0\nu\beta\beta$-decay amplitude ${\cal
T}_{0\nu\beta\beta}$ is given by~\cite{Klapdor}:
\begin{equation}
  \label{M2beta}
{\cal T}_{0\nu\beta\beta}\ =\ \frac{\langle m \rangle}{m_e}\ 
                         {\cal M}_{\rm GTF} (m_\nu )\; ,
\end{equation}
where ${\cal M}_{\rm  GTF} = {\cal M}_{\rm GT}  - {\cal M}_{\rm F}$ is
the  difference of  the   nuclear matrix  elements for   the so-called
Gamow-Teller  and  Fermi transitions.   Note   that this difference of
nuclear  matrix  elements  sensitively  depends   on the  mass  of the
exchanged KK neutrino in  a $0\nu\beta\beta$ decay (compare the discussion in
\cite{HK}). Especially if  the
exchanged KK-neutrino mass $m_{(n)}$ is comparable  or larger than the
characteristic  Fermi  nuclear   momentum $q_F  \approx 100$~MeV,  the
nuclear   matrix element    ${\cal     M}_{\rm  GTF}$  decreases    as
$1/m^2_{(n)}$.    The  general     expression  for    the    effective
Majorana-neutrino mass $\langle  m \rangle$ in~(\ref{M2beta}) is given
by
\begin{equation}
  \label{effMajmass}
\langle m \rangle\ =\ \frac{1}{{\cal M}_{\rm GTF} (m_\nu) }\
\sum_{n = -\infty}^{\infty}\, B^2_{e,n}\,m_{(n)}\, 
\Big[\, {\cal M}_{\rm GTF} (m_{(n)})\: 
-\: {\cal M}_{\rm GTF} (m_\nu )\, \Big]\; .\quad
\end{equation}
In the above, the  first term describes the genuine higher-dimensional
effect of KK-neutrino exchanges, while the second term is the standard
contribution  of  the  light   neutrino  $\nu$,  rewritten  by  virtue
of~(\ref{B2}).  Note that the dependence of the nuclear matrix element
${\cal  M}_{\rm GTF}$  on the  KK-neutrino masses  $m_{(n)}$  has been
allocated to  $\langle m  \rangle$ in (\ref{effMajmass}).   The latter
generally leads to predictions for  $\langle m \rangle$ that depend on
the  double beta  emitter isotope  used in  experiment.   However, the
difference in the predictions  is too small for the higher-dimensional
singlet-neutrino models  to be  able to operate  as a smoking  gun for
different $0\nu\beta\beta$-decay experiments.

To obtain  realistic predictions for the  double beta decay observable
$\langle m  \rangle$, 
we  have  used   the  general  formula~(\ref{effMajmass}),   where the
infinite sum  over $n$ has been truncated  at $|n_{\rm max}| = M_F R$,
namely at  the quantum gravity scale  $M_F$.  

%
%

\begin{table}[t]
\begin{center}
\begin{tabular}{|c|c|c|c|c|}
\hline
& \multicolumn{4}{c|}{}\\[-0.2cm]
$m_{(n)}$ [MeV] & \multicolumn{4}{c|}{${\cal M}_{\rm GTF} (m_{(n)})$} \\[0.2cm]
\hline
& & & & \\[-0.2cm]
& $^{76}$Ge & $^{82}$Se & $^{100}$Mo & $^{116}$Cd \\[0.2cm]
\hline
 $\le 1$ & $4.33$ &$4.03 $&$4.86$&$ 3.29$\\
 $10$ & $4.34$ &$4.04 $&$4.81$&$ 3.29$\\
 $10^{2}$ & $3.08$ &$2.82 $&$3.31$&$ 2.18$\\
 $10^{3}$ & $1.40\times 10^{-1}$ &$1.25\times 10^{-1} $&$1.60\times
10^{-1}$&$ 9.34\times 10^{-2}$\\
 $10^{4}$ & $1.39\times 10^{-3}$ &$1.24\times 10^{-3} $&$1.60\times
10^{-3}$&$ 9.26\times 10^{-4}$\\ 
 $10^{5}$ & $1.39\times 10^{-5}$ &$1.24\times 10^{-5} $&$1.60\times
10^{-5}$&$ 9.26\times 10^{-6}$\\ 
 $10^{6}$ & $1.39\times 10^{-7}$ &$1.24\times 10^{-7} $&$1.60\times
10^{-7}$&$ 9.26\times 10^{-8}$\\ 
 $10^{7}$ & $1.39\times 10^{-9}$ &$1.24\times 10^{-9} $&$1.60\times
10^{-9}$&$ 9.26\times 10^{-10}$\\[0.1cm] 
\hline
\hline
& \multicolumn{4}{c|}{}\\[-0.2cm]
$m_{(n)}$ [MeV] & \multicolumn{4}{c|}{${\cal M}_{\rm GTF} (m_{(n)})$}\\[0.2cm]
\hline
& & & & \\[-0.2cm]
& $^{128}$Te & $^{130}$Te & $^{136}$Xe & $^{150}$Nd \\[0.2cm]
\hline
 $\le 1$ & $4.50$&$3.89$&$1.83$&$5.30$\\
 $10$ & $4.52$&$3.91$&$1.88$&$5.45$\\
 $10^{2}$ & $3.19$&$2.79$&$1.48$&$4.24$\\
 $10^{3}$ & $1.46\times 10^{-1}$&$1.29\times 10^{-1}$&$7.07\times
10^{-2}$&$2.02\times 10^{-1}$\\  
 $10^{4}$ & $1.46\times 10^{-3}$&$1.28\times 10^{-3}$&$7.04\times
10^{-4}$&$2.02\times 10^{-3}$\\ 
 $10^{5}$ & $1.46\times 10^{-5}$&$1.28\times 10^{-5}$&$7.05\times
10^{-6}$&$2.02\times 10^{-5}$\\ 
 $10^{6}$ & $1.46\times 10^{-7}$&$1.28\times 10^{-7}$&$7.05\times
10^{-8}$&$2.02\times 10^{-7}$\\ 
 $10^{7}$ & $1.46\times 10^{-9}$&$1.28\times 10^{-9}$&$7.05\times
10^{-10}$&$2.02\times 10^{-9}$\\[0.1cm] 
\hline
\end{tabular}
\end{center}
\caption{\em QRPA estimates of the relevant combination of nuclear
matrix elements, ${\cal M}_{\rm GTF} = {\cal M}_{\rm GT}-{\cal M}_{\rm
F}$, as a function of the KK neutrino mass $m_{(n)}$.}\label{Tab1}
\end{table}

\begin{table}[t]
{\small 
\begin{center}
\begin{tabular}{|l|r|c|c|c|c|c|c|c|}
\hline
& \multicolumn{8}{c|}{}  \\[-0.2cm]
$1/a$  & \multicolumn{8}{c|}{$\langle m \rangle$~[eV]} \\[0.2cm] 
\hline
 & & & & & & & & \\[-0.2cm]
[GeV] & $^{76}$Ge & $^{82}$Se & $^{100}$Mo & $^{116}$Cd & $^{128}$Te & 
$^{130}$Te & $^{136}$Xe & $^{150}$Nd \\[0.2cm]
\hline
0.05  & 0.009 & 0.010 & 0.016 & 0.012 & 0.009 & 0.008 & --0.004  &
                                               --0.004 \\ 
0.1   & 0.052 & 0.054 & 0.061 & 0.062 & 0.052 & 0.050 & 0.025 & 
                                               0.026 \\ 
0.2   & 0.096  & 0.100 & 0.109 & 0.114 & 0.097 & 0.094 & 0.058 & 
                                              0.061 \\ 
0.3   & 0.123   & 0.128 & 0.136 & 0.143 & 0.124 & 0.121 & 0.082 & 
                                              0.086 \\ 
1     & 0.271 & 0.275 & 0.280 & 0.287 & 0.272 & 0.269 & 0.241 & 
                                           0.243 \\ 
10    & 0.493 & 0.493 & 0.494 & 0.495 & 0.493 & 0.493 & 0.489 & 0.489 
                                                  \\ 
\hline
10$^2$ & \multicolumn{8}{c|}{0.513} \\ 
10$^3$& \multicolumn{8}{c|}{0.535} \\ 
10$^4$& \multicolumn{8}{c|}{0.066} \\ 
10$^{10}$ & \multicolumn{8}{c|}{$\stackrel{<}{{}_\sim} 10^{-6}$}\\
\hline
\end{tabular}  
\end{center} }
\caption{\em Numerical estimates of $\langle m \rangle$ for different
nuclei in a 5-dimensional brane-shifted model, with $m=10$~eV, $1/R =
300$~eV, $\varepsilon = 1/(4R)$, $\phi_h = -\pi/4$ and $M_F = 1$~TeV.
}\label{Tab2}
\end{table}

In Table~\ref{Tab1}, we present numerical values for the difference of
the nuclear  matrix elements,  ${\cal  M}_{\rm  GTF} =  {\cal  M}_{\rm
GT}-{\cal   M}_{\rm F}$,  as   a function   of the   KK  neutrino mass
$m_{(n)}$.  Our estimates   are     obtained within  the     so-called
Quasi-particle         Random           Phase            Approximation
(QRPA)~\cite{Muto:cd,Staudt:qi}.   

In Table~\ref{Tab2},  we  show   numerical values for   the  effective
Majorana-neutrino  mass $\langle m \rangle$   as derived for different
nuclei in a 5-dimensional brane-shifted model,  with $m=10$~eV, $1/R =
300$~eV, $\varepsilon = 1/(4R)$, $\phi_h = -\pi/4$  and $M_F = 1$~TeV.
In addition, we have varied discretely  the brane-shifting scale $1/a$
from 0.05~GeV up to  values much larger  than $M_F$.  
It is obvious that,
within the extra-dimensional framework considered,
a sizeable  value for $\langle  m \rangle$  in the  presently explorable
range, is possible. 
Finally, for very  small values of $a$, i.e.\ for
$a \ll  1/M_F$, we obtain the  undetectably small result
for the unshifted brane $a=0$.

Apart from  explaining the  recent excess in  $0\nu\beta\beta$ decays,
the 5-dimensional model  with a small  but non-vanishing shifted brane
exhibits      another   very  important     property.   The  effective
Majorana-neutrino mass $\langle  m \rangle$ can  be  several orders of
magnitude larger than  the light neutrino   mass $m_\nu$, for  certain
choices of the parameters  $\varepsilon$ and $\phi_h$.  To  understand
this   phenomenon,   let      us  first    consider   the   eigenvalue
equation~(\ref{eigen}) for $\lambda = m_\nu$, written in the form:
\begin{equation}
  \label{B2approx}
m_\nu\: +\: \sum\limits_{n=-\infty}^\infty\,\frac{m^{(n) 2}}{
 \varepsilon\: +\: \frac{n}{R}\: -\: m_\nu  }\ =\ 0\; .
\end{equation}

The infinite  sum over KK neutrino states  can be performed with
the  help of~(\ref{transII}).  
Especially  for $a = \pi  R/q$ with $q$ being  an integer much
larger than  1, i.e.\ for  $1/M_F \ll a  \stackrel{<}{{}_\sim} 1/q_F$,
the light neutrino mass $m_\nu$ is given by
\begin{equation}
  \label{mnu}
m_\nu \ \approx\ -\, \pi m^2 R\ \Big[ \cos^2 \phi_h\, 
\cot (\pi R\,\varepsilon )\ +\ \frac{1}{2}\sin(2\phi_h)\, \Big]\, .
\end{equation}
It is now easy to see that the light neutrino mass $m_\nu$ can be very
suppressed  for  specific values  of $\phi_h$  and $\varepsilon$.  For
instance, one obvious choice   would  be $\phi_h \approx -\pi/4$   and
$\varepsilon \approx 1/(4R)$.   On the    other hand, the    effective
neutrino mass $\langle m \rangle$ is determined by the second
sine-dependent term in~(\ref{mnu}),  which  is
induced  by  brane-shifting  effects.    Unlike the  suppressed  light
neutrino mass $m_\nu$, the effective neutrino mass $\langle m \rangle$
can be sizeable in the  observable range of several tenths of an eV.  
This loss  of
correlation between the quantities  $\langle m \rangle$ and $m_\nu$ is
a  rather  unique feature   of  our   higher-dimensional brane-shifted
scenario. It provides an elegant mechanism for allowing the claimed
evidence for a non-zero  $0\nu\beta\beta$ signal~\cite{HMexp} 
to coexist with stringent
cosmological constraints on the absolute neutrino mass scale from the
cosmic microwave background and large scale structure surveys \cite{BPSW}.
As  discussed in \cite{Bhattacharyya:2002vf},  the  above
de-correlation property also plays a key  r\^ole  in model-building  of
5-dimensional brane-shifted scenarios that  could explain the neutrino
oscillation data.

\section{Conclusions}

We  have  studied  the  model-building constraints  derived  from  the
requirement  that KK singlet  neutrinos in  theories with  large extra
dimensions can  give rise to a  sizeable $0\nu\beta\beta$-decay signal
\cite{Bhattacharyya:2002vf}.   
Our  analysis has  been
focused on  5-dimensional $S^1/Z_2$  orbifold models with  one sterile
(singlet) neutrino in the bulk,  while the SM fields are considered to
be  localized on  a  3-brane.   In our  model-building,  we have  also
allowed the 3-brane to be  displaced from the $S^1/Z_2$ orbifold fixed
points.   Within this  minimal 5-dimensional  brane-shifted framework,
lepton-number  violation  can   be  introduced  through  Majorana-like
bilinears,  which  may  or  may  not arise  from  the  Scherk--Schwarz
mechanism,  and   through  lepton-number-violating  Yukawa  couplings.
However, lepton-number-violating  Yukawa couplings can  be admitted in
the theory, only if the 3-brane is shifted from the $S^1/Z_2$ orbifold
fixed  points.   Apart   from  a  possible  stringy  origin~\cite{GP},
brane-shifting might also be regarded  as an effective result owing to
a  non-trivial 5-dimensional profile  of the  Higgs particle~\cite{RS}
and/or other SM fields~\cite{AHS,KKS} that live in different locations
of a 3-brane  with non-zero thickness which is centered  at one of the
$S^1/Z_2$ orbifold fixed points.

One   major difficulty of the    higher-dimensional theories is  their
generic   prediction  of a   KK  neutrino  spectrum  of  approximately
degenerate states with opposite  CP parities that lead to  exceedingly
suppressed values for the  effective Majorana-neutrino mass $\langle m
\rangle$.  Nevertheless, we have  shown that within the  5-dimensional
brane-shifted framework, the  KK neutrinos can  couple  to the $W^\pm$
bosons with unequal  strength, thus avoiding the  disastrous CP-parity
cancellations in the  $0\nu\beta\beta$-decay amplitude. In particular,
the     brane-shifting parameter  $a$   can   be  determined from  the
requirement that the effective Majorana mass $\langle m \rangle$ is in
the observable range. In this way, we have
found  that  $1/a$ has  to be  larger  than the  typical Fermi nuclear
momentum $q_F  = 100~{\rm  MeV}$  and much  smaller  than the  quantum
gravity     scale    $M_F$,   or     equivalently    $1/M_F    \ll   a
\stackrel{<}{{}_\sim} 1/q_F$.

An important  prediction of our  5-dimensional  brane-shifted model is
that the effective Majorana-neutrino mass  $\langle m \rangle$ and the
scale of light neutrino masses   can be completely de-correlated   for
certain    natural  choices  of     the   Majorana-like bilinear  term
$\varepsilon$ and the  original 5-dimensional Yukawa couplings $h^l_1$
and  $h^l_2$ in~(\ref{Leff}).   For example,  if  $\varepsilon \approx
1/(4R)$ and  $h^l_1 \approx -  h^l_2$, we obtain light-neutrino masses
that   can be several  orders  of  magnitude smaller  than $\langle  m
\rangle$.

\end{document}